\documentstyle[12pt]{article}

\begin{document}

\def\hs{\hskip}
\def\vs{\vskip}
\def\ni{\noindent}
\def\pa{\parindent}
\def\i{\item}
\def\ii{\itemitem}
\def\ej{\vfill\eject}
\def\hf{\hfill}
\def\k{$\vec k_1$ }
\def\kk{$\vec k_2$ }
\def\p{$\vec p_1$ }
\def\pp{$\vec p_2$ }
\def\kp{\vec k_i,\vec p_i }
\def\to{${\cal T}_0$}

\baselineskip=23pt

\rightline {OITS-592}
\rightline {November 1995}

\vskip2cm

\centerline{\large\bf CHARM CORRELATION AS A DIAGNOSTIC PROBE}

\vskip.5cm
\centerline{\large\bf OF QUARK MATTER}

\vskip1.5cm
\centerline{\large\bf Rudolph C. Hwa}
\vskip.5cm
\centerline{Institute of Theoretical Science and Department of Physics}
\centerline{University of Oregon, Eugene, OR 97403} \vskip2cm
\centerline{\bf Abstract}
{\begin{quotation}
The use of correlation between two open-charm mesons is suggested to
give information about the nature of the medium created in heavy-ion
collisions.  Insensitivity to the charm production rate is achieved by
measuring normalized cumulant.  The acollinearity of the $D$ momenta
in the transverse plane is a measure of the medium effect.  Its
dependence on nuclear size or $E_T$ provides a signature for the
formation of quark matter.

\end{quotation}}
\vfill\newpage
        The conventional probes of quark matter in heavy-ion collisions, such
as dileptons and $J/\psi$ suppression, have not provided conclusive
evidence about the creation or absence of quark-gluon plasma
\cite{bm}.  A large part of the difficulties involve the ambiguities
arising from competing processes and from uncertainties in the initial
normalizations of some key quantities.  An effective probe should be
free of such ambiguities.  In this paper we suggest the possibility
that charm correlation may be such a probe.

        The correlation proposed is between open-charm mesons at nearly
opposite directions.  The charm quark is used both for tagging and
for probing.  Heavier quark can be used when appropriate; charm will
be used as a generic term for heavy quark in the following
discussion.  Lighter partons and associated minijets are too
copiously produced at RHIC and LHC \cite{kk} to be useful for our
purpose of tagging and probing.  The idea is based simply on the dual
requirements that the signature should be independent of the
production rate but sensitive to the medium through which the probe
traverses.  Appropriately normalized cumulant can satisfy the first
requirement, while the transverse deviation from exact back-to-back
correlation meets the second.

Briefly stated, it is suggested that one searches for
$D\bar{D}$ produced in the transverse plane at $y=0$ with their
momenta nearly collinear, but opposite.  The acollinearity in the
transverse plane is the measure of interest.  To enhance the effect,
experimental cuts should be made on the magnitudes of the $D$-meson
momenta so that they are nearly equal and not too large.  We expect
that the mean acollinearity is smaller if the medium is deconfined quark
matter than if it is not.

The proposed measure is similar in spirit to the acoplanarity of
jets suggested by Appel \cite{da}, but significantly different in
substance.  The major differences are:  (a) the jet axes cannot be
as precisely determined as the $D$-meson momenta, (b) at high
energy too many jets are produced resulting in contamination and
deterioration of the correlation signal, (c) we emphasize the
difference between the propagation of a $c$ quark through a
deconfined medium and that of a $D$-meson through a confined
medium, and (d) the phenomenology of $D\bar{D}$ correlation
can reveal interesting physics even in kinematic regions where
perturbative QCD (pQCD) is unreliable.

Let \p and \pp be the momenta of the two detected
charge-conjugate $D$ mesons.  The cumulant is
\begin{eqnarray}
c(\vec p_1,\vec p_2) = \rho_2(\vec p_1,\vec p_2) - \rho_1(\vec
p_1)\rho_1(\vec p_2),
\label{1}
\end{eqnarray}
where $\rho_n$ is the $n$-particle distribution function; it is
the irreducible part of the two-particle correlation.  In a
heavy-ion collision the cumulant can in general be expressed in the
form
\begin{eqnarray}
c(\vec p_1,\vec p_2) =\int {d^3k_1 \over k^0_1} {d^3k_2 \over k^0_2}
S(\vec k_1,\vec k_2)H(\vec k_1,\vec p_1)H(\vec k_2,\vec p_2),
\label{2}
\end{eqnarray}
where $S(\vec k_1,\vec k_2)$ is the probability of producing
two partons with momenta \k and \kk, and $H(\vec k_i,\vec p_i)$ is
the hadronization function that connects the parton $i$ at the point of
creation to the hadron detected with momentum $\vec p_i$.  Hereafter we
adopt the convention of using the symbol $k\ (p)$ for parton
(hadron) momentum.  It should be stressed that $H$ is not
simply the fragmentation function usually used for jet considerations
because firstly the produced parton must traverse a dense medium and
suffer momentum degradation before fragmentation, and secondly the
hadronization process may be recombination \cite{kd,rh} instead of
fragmentation.  In fact, it has been shown that the data of open charm
production in the forward region of hadronic collisions can be well
described by recombination \cite{rh2}, but badly by pQCD or
fragmentation model \cite{ga}.

In the domain where pQCD is reliable one can write $S(\vec k_1,\vec
k_2)$ for $AB$ collision as
 \begin{eqnarray}
S(\vec k_1,\vec k_2) =c \int {d^3k_a \over k^0_a} {d^3k_b \over k^0_b}
F_A (k_a) F_B(k_b) \delta^4 (k_a+k_b-k_1-k_2)
\left|M(a+b\rightarrow1+2)\right|^2
\label{3}
\end{eqnarray}
plus other terms of similar structure, if more than one hard
subprocess are important.  In (\ref{3}) $c$ is a numerical constant,
$F_A(k_a)$ is the parton distribution in nucleus $A$, and
$M(a+b\rightarrow1+2)$ is the amplitude of the hard subprocess
involved.  There is a great deal of physics contained in the
determination of $F_{A,B}$, which depends on nucleon structure function
at small $x$, gluon distribution, nuclear shadowing, initial-state
radiation, preequilibrium and possibly thermal
interactions, space-time evolution, etc.  So much uncertainty is involved
in the problem that the study of open-charm production has been
suggested as a means to learn more about the parton dynamics in the
early phase of nuclear collision \cite{zl,pl}, i.e., the reverse of using
$F_{A,B}$ to predict measureable quantities.  While that is certainly a
worthwhile project to pursue, our proposal here is to circumvent all
that complication and proceed with the use of $S(\vec k_1,\vec k_2)$
independent of the details about $F_{A,B}$.

{}From $S(\vec k_1,\vec k_2)$ not only can two-particle inclusive
distribution be determined as in (\ref{2}), the one-particle
distribution can also be obtained as follows
\begin{eqnarray}
\rho_1 (\vec p_2) = \int {d^3 k_1 \over k^0_1} {d^3k_2 \over k^0_2}
S(\vec k_1,\vec k_2) H(\vec k_2,\vec p_2) .
\label{4}
\end{eqnarray}
To free our signal from the uncertainties of the primordial parton
dynamics, let us define the singly-normalized cumulant function
 \begin{eqnarray}
C(\vec p_1,\vec p_2) = c(\vec p_1,\vec p_2)/ \rho_1 (\vec p_2).
\label{5}
\end{eqnarray}
It is clear from (\ref{2}) and (\ref{4}) that $C(\vec p_1,\vec p_2)$
should be insensitive to the rate of charm production.

Hereafter we shall regard hadron 2 (with momentum $\vec p_2$) as the
trigger particle, against which we study the properties of the probe
particle 1.  Of course, it is the relative momentum between the
trigger and probe that is important, but conceptually it is efficient
to identify (arbitrarily) one of the two $D$ mesons as the trigger
and define the axes such that the trigger momentum in every event is
always aligned along a fixed direction, say $-\hat x$ axis, with
$\hat y$ being the axis normal to the scattering plane containing the
beams and the trigger.  The aim is to study the momentum
distribution of the other $D$ meson in the neighborhood of the
$+\hat x$ axis.

Since the parton momenta \k and \kk can have large longitudinal
imbalance due to unequal momenta, $\vec k_a$ and $\vec k_b$, of the
initial colliding partons, but they can have only limited total
transverse momentum $\vec K_T = \vec k_{1_T} + \vec k_{2_T}$ due to the
small intrinsic transverse momenta of the initial partons and to the
initial-state radiation, we can avoid the complexity of the full
structure of $S(\vec k_1,\vec k_2)$ if we restrict \k and \kk to only
the near neighborhood of a common transverse plane.  That is achieved
by requiring that \p and \pp lie only in the transverse plane at $y=0$.
For brevity we shall refer to that plane as \to.  Since we expect the
angular differences between $\vec k_i$ and $\vec p_i$ to be small, that
requirement therefore forces
$\vec k_1$ and $\vec k_2$ to be very close to \to \ also.

The aim of this problem is to learn about the medium effect through
$H(\vec k_i,\vec p_i)$, on which the measureable $C(\vec p_1,\vec p_2)$
depends.  With $\vec k_i$ near \to \ those partons do not
participate in the longitudinal expansion of the system, which is
another area of large uncertainties.  But even in \to \ there are
several possible processes leading from the partons $(k_i)$ to the
hadrons $(p_i)$, each involving a different hadronization function
$H(\vec k_i,\vec p_i)$.  So far we have not specified the kinds of
partons carrying $k_i$.  They can be high-momentum quarks of the
$u$ and $d$ types, or lower-momentum gluons, all capable of
fragmenting into the $D$ mesons.  Since the two fragmentation
processes are independent, $\vec p_1$ and $\vec p_2$ are not
correlated even though $\vec k_1$ and $\vec k_2$ are.  To narrow
down the hadronization process we make use of the experimental
freedom to require further that the magnitudes $p_1$ and $p_2$ are
nearly equal within a narrow range.  Moreover, that magnitude should not
be too high, say, in the $2-5$ GeV range.  In so doing we can maximize
the contribution from $c \bar {c}$ pair creation to the formation
of $D \bar {D}$.  There are
several stages of reasoning involved here, which we now describe.

The possible parton types are $q$ and $c$, where $q$ denotes $u$,
$d$, or $g$ collectively.  Let the flavor labeling not be encumbered
by concerns about quark or antiquark differences.  We postpone our
consideration about the $s$ quark until later.  The possible
hadronization processes are fragmentation $(F)$ and recombination
$(R)$, for which $k>p$ in $F$, but $k<p$ in $R$.  Thus for the
production of $D$ there are four possible processes:  $F(q
\rightarrow D)$, $F(c \rightarrow D)$, $R(q \rightarrow D)$, and $R(c
\rightarrow D)$.  At high collision energies there are so many hard
subprocesses \cite{kk} that there are enough transversly moving
partons to make recombination competitive with fragmentation in the
formation of $D$ in \to.  That is not the case in $pp$ or $pA$
collisions.  In fact, for any given momentum $p$ of $D$ in \to, it is
more favored to recombine two lower momenta partons to add up to $p$
than to create a higher momentum parton which subsequently decays to
$p$, since the probability of creating high $k_T$ partons falls off
as a power $k_T^{-\alpha}$, with $\alpha >4$.  For single-particle
inclusive distribution a comparison between the two hadronization
processes in the production of particles in \to \ has been studied
quantitatively \cite{rh} with the result that for $p_T<6$ GeV/c and for
large nuclei the rate of hadronization through recombination is at least
an order of magnitude higher than through fragmentation.  For
a pair of correlated particles the $R/F$ ratio of the rates would be
squared.  Thus we may ignore $F(q \rightarrow D)$ and $F(c \rightarrow
D)$ in the following discussion.

Since $m_c \approx 5\,m_q$ for constituent quark masses, the momentum
fractions $x_c$ and $x_q$ of $c$ and
$q$, respectively, in $D$ are on the average very different, with $x_c
\approx 5\,x_q$; hence, the recombination function for $c+q \rightarrow
D$ is maximum when $\vec k_c\approx 5\,\vec k_q$ in the same direction
\cite{rh2,rh3}.  For \p  and \pp  nearly equal and opposite, the
production of
$D \bar{D}$ is therefore dominated by the creation first of $c
\bar{c}$ pair with $\vec k_1\approx -\vec k_2$ followed by
recombination with low-momentum $q$ quarks, rather than by the process
where a created $q \bar{q}$ pair dictates the momenta of the $D$
mesons.  In short, $R(c \rightarrow D)$ is more important than $R(q
\rightarrow D)$.  In the following we shall focus on the process where
$S(\vec k_1, \vec k_2)$ in (\ref{2}) describes the hard production of a
$c \bar{c}$ pair, and the two $H$ functions represent $R(c
\rightarrow D)$.

The only part in the problem that has a firm theoretical footing is
the amplitude $M$ for hard scattering in (\ref{2}), which is calculable
in
pQCD.  Even there, charm production with $k_T \sim 2$ GeV/c is in
the grey area of reliability.  As mentioned earlier, the parton
distribution $F_A(x_a)$ and $F_B(x_b)$ are quite uncertain, but the
normalized cumulant $C(\vec p_1, \vec p_2)$ is insensitive to all of
them, including $M$; furthermore, we emphasize the misalignment of
\p  and \pp, their magnitudes being selected by experimental cuts.
Thus the signature we seek depends mainly on $H(k_i \rightarrow
p_i)$, which is sensitive to the medium that stands between the
creation of $c \bar{c}$ and the detected hadrons $D
\bar{D}$.  That is just what a good probe should be.  It is
unfortunate that $H(k_i \rightarrow p_i)$ cannot at this point be
calculated precisely in QCD, perturbative or otherwise.  However, the
discovery of unambiguous experimental signature is more important than
having reliable theoretical calculations at this stage.  On the basis
of reasonable arguments we indicate below what that signature might
look like.

The four vectors $\vec k_i$ and $\vec p_i$ $(i=1,2)$ are all very
close to \to.  We consider below only their projections $\vec k_{i_T}$
and $\vec p_{i_T}$ on \to.  For brevity we shall omit the subscripts
$T$, unless there is confusion.  Let the angles among these four
vectors be labeled as follows:  $\theta_1 (\vec k_1, \vec p_1)$,
$\theta_2 (\vec k_2, \vec p_2)$, $\theta_{12} (\vec k_1, -\vec k_2)$,
and $\phi (\vec p_1, -\vec p_2)$.  For simplicity let the
distributions in these angles be represented by Gaussians:  ${\rm
exp} (-\theta_1^2/ \lambda_1^2)$, ${\rm exp} (-\theta_2/
\lambda_2^2)$, ${\rm exp} (-\theta_{12}^2/ \lambda_{12}^2)$, and ${\rm
exp} (-\phi^2/ \lambda^2)$, respectively.  The distribution in
$\phi$ is a Gaussian also because of the convolution theorem, since
(\ref{2}) implies
\begin{eqnarray}
c(\vec p_1, \vec p_2) \propto \int d\theta_1 d\theta_2\,{\rm exp}
 \left( - {\theta_1^2 \over \lambda_1^2}
-{\theta_{12}^2 \over \lambda_{12}^2}-
{\theta_2^2 \over \lambda_2^2}\right) ,
\label{6}
\end{eqnarray}
where $\theta_{12}= \phi - \theta_1 - \theta_2$.  Thus if $\lambda_1$,
$\lambda_2$ and $\lambda_{12}$ are all small compared to $\pi$, then
\begin{eqnarray}
\lambda^2 = \lambda_1^2 + \lambda_2^2 + \lambda_{12}^2  .
\label{7}
\end{eqnarray}
One can improve on these integrations with better knowledge on
$S(\vec k_1, \vec k_2)$ and $H(\vec k_i, \vec p_i)$, but the result
will not differ much from (\ref {7}), which encapsules the property that
the width in $\vec p_1\cdot$\pp  is the rms sum of those in $\vec
p_1\cdot$\k,
$\vec k_1\cdot$\kk  and $\vec p_2\cdot$\kk.  Since we require $|\vec p_1|
\approx |\vec p_2| \equiv p$, we may write $C(\vec p_1, \vec p_2)$ as
$C(p, \phi)$.  Then we have
\begin{eqnarray}
C(p, \phi) = C(p)\,e^{-\phi^2/ \lambda^2} .
\label{8}
\end{eqnarray}
Alternatively, one can define $\vec p_1=-\vec p_2+\vec p_t$ (note the
small t), where $\vec p_t$ is nearly normal to the trigger axis but
still in
\to.  Thus experimentally, the data on $C(\vec p_1, \vec p_2)$ can be
presented as distributions in $\xi \equiv p_t/p$ for various values of
the trigger momentum $p$.  Empirical $C(p, \xi)$ should be sharply
peaked in $\xi$, though not necessarily Gaussian; the width can be
represented by $\lambda$ in (\ref{8}) in the following discussion.

In order to keep all the quantities in (\ref{7}) small, $p$ should not
be too small; neither should it be too large so that the rate of
producing two correlated $D$ mesons does not become too low.  The
range $2<p<5$ GeV/c appears to be reasonable.  In that range the $c$
and $\bar{c}$ quarks are beams in the dense medium with momenta
$k_1$ and $k_2$, both $< p$, insufficient to give any hope to the
validity of pQCD in describing their passages through the medium.
Nevertheless, we expect charm quarks in that momentum range to be
sensitive to the medium effects and can provide us with useful
information.

We now consider the various contributions to (\ref{6}).  Nonzero
$\theta_{12}$ means that $\vec K \equiv \vec k_1 + \vec k_2 \neq 0$.
Nonzero $\vec K$ is due to the intrinsic transverse momenta of the
partons plus the recoil from initial-state radiation before the hard
subprocess that creates the $c \bar{c}$ pair.  We expect $|\vec
K|$ to be small, and set its average value at $0.3$ GeV/c for low
$k_i$ and allow it to be higher for higher $k_i$.  Approximating
$k_1=k_2$ and denoting them collectively as $k$, we have
cos\,$\theta_{12} =1-K^2/2k^2$, and $\theta_{12} \simeq K/k$.  Thus for
$k>2$ GeV/c, we have $\lambda_{12} \leq 0.15$.  More importantly,
$\lambda_{12}$ is independent of the medium in which the created $c$
quarks will traverse, and will therefore not affect our signature of
the medium effect.

Next, we consider the last two terms of (\ref{6}) due to the $H$
functions in (\ref{2}), which are the heart of the problem.  The two
$H$ functions independently describe the hadronization processes from
$\vec k_i$ to $\vec p_i$, as they proceed along the paths $l_1$
and $l_2$, through the medium in essentially opposite directions in
\to  \ from the point of creation of $c \bar{c}$.  Since the result we
seek concerns the sum $\lambda_1^2 + \lambda_2^2$ in (\ref{7}), it is
equivalent to that due to one $c$ quark traversing the entire path
$L= l_1+ l_2$ from one end of the medium to the other in \to, passing
through the point of $c \bar{c}$ creation.  The transverse
expansion of the medium is not rapid.  If $R$ is the average radius
of the system that is relevant to this study, then when averaged over
all possible points of the $c \bar{c}$ creation in \to \ and over
all orientations of the path, the mean path length $L$ of crossing a
disc of radius $R$ is $L= 4R/3$.  Our consideration is thus reduced
to the hadronization of a $c$ quark with initial momentum $\vec k$
passing through a slab of the medium of thickness $L$ and emerging
as a $D$ meson with momentum $\vec p$.

There are two scenarios to examine.  One is that the medium consists
of deconfined quarks and gluons, while the other is of high-density
hadrons with quarks confined.  Let them be referred to as quark
matter (QM) and hadron matter (HM), respectively.  Of course, they
represent two extreme cases, and other scenarios that stand between
them are possible.  If we know the nature of the signatures for the
extreme cases, what happens in the intermediate cases can be
estimated by interpolation.  Thus for now it is sensible to consider
just QM and HM.
 \vskip.3cm
\noindent 1.  {\it Quark Matter}.\quad  Being deconfined, the medium
cannot support the formation of any hadrons nor the existence of any
color flux tubes in it.  Thus the $c$ quark that traverses the QM
medium remains as a $c$ quark.  It may lose momentum and deviate from
straightline path, but the formation of $D$ can occur only after
passing through the medium by recombining with a $\bar{q}$ at
the exit point.  If pQCD were applicable, one could study in detail
the effect of multiple scattering as in \cite{xw,mt}, and
determine the degree of energy loss and $k_t$ gain.  Hereafter we use
$k_t$ to denote parton momentum transverse to $\vec k$.  However, for
$k$ as low as $2$ GeV/c, the reliability of pQCD is questionable.
Qualitatively, one expects the radiative energy loss to be reduced
for heavy quarks compared to light quarks.  We assume that the loss
of longitudinal momentum is roughly compensated by the gain in
momentum due to recombination so that $p \approx k$.  The more
important aspect of the problem is the $k_t$ gain. If one regards the
result of \cite{xw} that takes the LPM effect into account as being
valid, then $k_t$ due to gluon radiation is of the order of the color
screening mass $\mu$, independent of the number of multiple
scatterings.  The cumulative effect on $k_t$ due to elastic scatterings
depends on whether the random-walk model is valid or the quantum
coherence effect is important.  In the former case the process is
Markovian and $k_t$  would increase with $L$, while in the latter case
it would not.  The coherent LPM effect in the longitudinal component
is non-Markovian. It has been known that the former is more relevant to
conventional large-$p_T$ processes, while the latter is for low-$p_T$
processes.  Our $H(c\rightarrow D)$ in the present problem belongs to
the latter category.  The following experimental facts support  the
latter.

Exhaustive studies of $h_1 A \rightarrow h_2X$ inclusive reactions at
high energies and low $p_T$ have revealed that, for $h_2$ in the beam
fragmentation region of $h_1$, the $p_T$ distribution of $h_2$ is
essentially independent of $A$ \cite{ps,db}.  Because of the flavor
dependence of the $h_2$ distribution, the processes can be well
interpreted in the parton picture by considering the valence quark
that is common in $h_1$ and $h_2$ \cite{kd,rh3,lg}.  Take
$h_1=\pi^+$ and $h_2=K^+$ to be specific.  For $K^+$ with high
$x_F$, it is the high momentum $u$ quark in $\pi^+$ that leads to
$K^+$ by recombination.  The $p_T$ of $K^+$ reflects the $k_T$ of the
$u$ quark after traversing the target.  The independence of
$\langle p_T\rangle$ on $A$ implies that there is no $k_T$ broadening due
to multiple-scattering effect on the $u$ quark.  Specifically, the
data of \cite{db} indicate that $\sigma (p_T=0.5)/\sigma(p_T=0.3)$
stays essentially uncharged at 0.5 and 0.48 for $A=Cu$ and $Pb$,
respectively, for $p_{\pi^+}=100$ GeV/c  and $p_{K^+}=80$ GeV/c.
Although it is a $u$ quark traversing normal nuclei, whereas our
problem involves a $c$ quark traversing dense QM, the independence
of $k_t$ on the path length is likely to be due to a common origin:
quantum coherence in low-$k_t$ processes is non-Markovian, so $k_t$ does
not increase with $L$.  The same lack of $A$ dependence is found in
\cite{ps} for
$h_1=p$ and $h_2=\Lambda$.  We emphasize that it is true only at low
$p_T$.  Significant $A$ dependence at high $p_T$ is not excluded, such
as in the production of massive dileptons in $pA$
collisions \cite{bor, ald}, where no $A$ dependence is seen until $p_T$
exceeds 2 GeV/c.

On the basis of these arguments we adopt the following position.
Firstly, the magnitude of $k_t$ is small, since heavy quark suffers
less deflection; we take it to be of order $\mu$, which according to
$\mu^2= 4\pi \alpha_s T^2$ gives $\mu \approx 0.4$ GeV for
$\alpha_s=0.3$ and $T= 200$ MeV.  Secondly, $k_t$ is expected to be
independent of $L$, although a gentle increase with $L$ cannot be
ruled out on firm theoretical ground.  Since the recombination of $c$
with a $\bar {q}$ to form a $D$ after passing through the QM
does not increase $p_t$ beyond the $k_t$ gained, we arrive at the
result that $\lambda_1^2 + \lambda_2^2 \approx \lambda_{12}^2$.
Consequently, we have $\lambda \approx 0.5/p$, where $p$ is in units
of GeV/c.  It is clear that if we want $\lambda$ to be small, $p$
should not be small.  That is why we have set $p>2$ GeV/c.
 \vskip.3cm
\noindent
2.  {\it Hadron Matter}.\quad  Consider now the scenario where
the created $c \bar{c}$ pair find themselves in a densly packed
hadronic medium.  The recombination of $c$ with $\bar{q}$ and
$\bar{c}$ with $q$ take place rapidly, and it is the $D$ and
$\bar{D}$ that traverse the HM in opposite directions.  As
before, the combined effect on \p  and \pp  can be represented by a $D$
of momentum $p$ going through a slab of HM of thickness $L$.  Being a
low-momentum hadron ($2<p<5$ GeV/c) the $D$ interacts strongly with
the mainly pionic medium.  The number of multiple collisions in $L$
is $\nu =n_\pi \sigma_{D\pi} L$, where $n_\pi$ is the pion density.
The energy dependence of $\sigma_{D\pi}$ is not known, but its
magnitude (in the few mb range) is definitely much greater than
partonic cross section.  Since $\lambda_1^2 + \lambda_2^2$ is
proportional to $\nu$, we therefore expect $p_t$ to be larger
(compared to the QM case) and to increase significantly with $L$.
Herein lies the major difference between the two medium effects on
$\xi =p_t/p$. Since $\lambda_{12}$ is the same for the two media, the
difference does not depend on it.
 \vskip.3cm
Putting together the above considerations leads to our suggestion
for the signature of QM vs. HM.  Measure $C(p,\xi)$.  Plot the $\xi$
dependence for fixed $p$ and determine the mean $\bar{\xi}$.
Examine how $\bar{\xi}$ depends on $E_T$ and $A$.  For HM,
$\bar{\xi}$ should be large and increase with $E_T$ and $A$.
If, at high enough $s$, $E_T$ and $A$, quark matter is created,
then $\bar{\xi}$ should drop down to a low value and become
essentially independent of $E_T$ and $A$.  This transition from
high and increasing $\bar{\xi}$ to low and roughly constant
$\bar{\xi}$ is the signature of QM formation.  In short, quarks are
smaller than mesons; their difference should be revealed in the
measurement of $\bar\xi$.

It would also
be of interest to study the $p$ dependence of $\bar{\xi} (p)$
.  If $\sigma_{D\pi}$ increase with $p$, then $\bar{\xi}_{\rm HM}$
would be constant in $p$, in contradistinction from
$\bar{\xi}_{\rm QM}(p)$ which decreases with $p$.

Another interesting problem arises when the recombination of $c\
(\bar{c})$ with $\bar{s}\ (s)$ is considered.  Since in QM
the hadronization occurs outside the quark phase, the width
$\bar{\xi}_{D_s \bar{D}_s}$ for $D_s \bar{D}_s$
correlation should not differ significantly from $\bar{\xi}_{D
\bar{D}}$.  However, in HM the formation of $D_s$ and
$\bar{D}_s$ occurs inside the medium, and because of the
suppressed $D_s
\pi$ scattering due to the OZI rule at low energy, we expect
$\bar{\xi}_{D_s
\bar{D}_s}<\bar{\xi}_{D \bar{D}}$.  The observation
of these differences would add to our understanding of what occurs
in these systems.

What is described in this paper is for idealized systems.  In
reality the system may be much more complicated and the
transition of the behavior of $\bar{\xi}$ may be very
gradual.  If so, it is unlikely that any other signature would be
clear-cut, since it is the system itself that is not sufficiently
distinctive.  Our proposal deals with the nature of the matter
probed, independent of other inessential complications, such as
the absolute normalization of the charm production rate, the
precise value of the initial temperature, or the validity of
pQCD.  To have a proper theoretical treatment of the problem is
essential ultimately, but for now the need for a distinctive
experimental signature seems to be more urgent.  Even if the
suggested signature turns out to be ineffective because of the
complexity not considered in this initial investigation, charm
correlation should nevertheless reveal much information about
heavy-ion collisions not available so far.

I am grateful to X.N. Wang for helpful discussions.  This work was
supported, in part, by the U.S. Department of Energy under grant No.
DE-FG06-91ER40637.

\end{document}